**Atomic-Scale Light Coupling Control in Ultrathin Photonic Membranes**

*Chih-Zong Deng, Chun-Hao Chiang, Sunhao Shih, Jui-Han Fu*, Yen-Ju Wu, Kuniaki Konishi, Vincent Tung, Chun-Wei Chen, and Ya-Lun Ho[1]**

C.-Z. Deng, C.-H. Chiang, Y.-L. Ho
Research Center for Electronic and Optical Materials, National Institute for Materials Science (NIMS), Ibaraki, Japan
E-mail: HO.Ya-Lun@nims.go.jp

S. Shih, J.-H. Fan, V. Tung
Department of Chemical System Engineering, School of Engineering, The University of Tokyo, Tokyo, Japan

Y.-J. Wu
Center for Basic Research on Materials, National Institute for Materials Science (NIMS), Ibaraki, Japan

K. Konishi
Institute for Photon Science and Technology, School of Science, The University of Tokyo, Tokyo, Japan

C.-W. Chen
Department of Materials Science and Engineering; Center of Atomic Initiative for New Materials, National Taiwan University, Taipei, Taiwan

Funding: JSPS KAKENHI Grant Number JP23K26155, JP21H04660; Advanced Research Infrastructure for Materials and Nanotechnology in Japan (ARIM) of the Ministry of Education, Culture, Sports, Science and Technology (MEXT), Japan (Proposal Number JPMXP1225NM5090); Areas Research Center Program, Higher Education Sprout Project by the Ministry of Education in Taiwan (Grant Number. 111L900801).

**Abstract text.**

Atomic-layer and two-dimensional (2D) materials have emerged as essential building blocks for next-generation quantum and semiconductor technologies, where atomic-scale control over light-matter interactions is critical. However, their inherently small interaction volume poses fundamental challenges for efficient integration into quantum and nanophotonic devices. Addressing this limitation requires the development of photonic platforms that can effectively enhance atomic-scale optical coupling. To this end, freestanding nanomembranes with extreme thinness and minimal radiative loss offer an ideal framework for integrating these materials into photonic systems. Here, we demonstrate an ultrathin photonic nanomembrane enabling atomic-scale control of light coupling. This architecture supports strong field confinement at the surface and significantly enhances light-matter interaction. Through the integration of atomic-layer dielectrics, we achieve Å-level thickness modulation, where each deposition cycle leads to an ultrafine shift of the high-Q resonance. High-resolution spatial mapping further confirms uniform and deterministic resonance tuning across the nanomembrane surface. Furthermore, by integrating a $WS_2$ monolayer with the photonic nanomembrane, strong field localization within the monolayer and a significant emission enhancement are achieved. This approach offers a scalable and versatile route for atomic-scale light coupling, helping to overcome the limitations of conventional photonics and opening opportunities in quantum photonics, optoelectronics, and advanced semiconductor technologies.

## 1. Introduction

Atomic-layer materials, including two-dimensional (2D) materials and atomically precise dielectric films grown by atomic layer deposition (ALD), with thicknesses ranging from a few angstroms to several nanometers, have emerged as foundational components in nanophotonics, optoelectronics, quantum technology, and advanced semiconductor research.[1–15] Their extreme dimensional confinement enables significant light-matter interactions and offers unique opportunities for manipulating light at the atomic scale. However, the inherently small optical interaction volume of these materials imposes critical challenges on their integration into photonic systems, as their extreme thinness limits their ability to confine optical fields or effectively modulate resonant optical modes. To fully exploit their potential, it is essential to design nanophotonic platforms that not only provide field confinement at the interface with the atomic-layer materials but also exhibit dimensional compatibility with these atomically thin materials to ensure efficient coupling and strong near-field interaction.

However, conventional nanophotonic architectures are typically orders of magnitude thicker than atomic-layer materials, leading to substantial mode mismatch and inefficient coupling.[16–20] Furthermore, many nanophotonic platforms rely on substrate-based structures, which suffer from inherent limitations such as substrate-induced optical loss, constrained modal confinement, and resonance rigidity with minimal spectral tunability, all of which hinder effective interaction with atomically thin materials. This challenge is particularly pronounced for atomic-layer materials, where extreme thickness constraints impose stringent requirements on photonic designs capable of sustaining high-Q optical modes while preserving near-field enhancement. In addition to integration challenges, characterizing atomically thin materials remains difficult due to their minimal thickness and the limitations of conventional techniques.[21] While transmission electron microscopy (TEM) offers atomic-scale resolution, it requires high-vacuum conditions and extensive sample preparation, and it can potentially damage fragile structures. Non-destructive methods such as X-ray reflectometry (XRR) and spectroscopic ellipsometry (SE) demand large, uniform areas and are sensitive to interface roughness, reducing accuracy for ultrathin films. Atomic force microscopy (AFM) provides high spatial resolution but only probes localized regions and lacks direct optical sensitivity. These constraints highlight the need for non-invasive, high-sensitivity optical methods that are compatible with ambient conditions and suitable for probing atomic-layer materials.

Among various nanophotonic approaches, platforms engineered to support high-Q modes–particularly bound states in the continuum (BICs)–offer unique advantages for light confinement and interaction enhancement. BICs arise from destructive interference between radiative channels, enabling optical modes to exhibit minimal radiative losses and achieve extreme field confinement with theoretically infinite Q-factors.[22,23] Like many nanophotonic structures, many BIC-based implementations rely on substrate-supported configurations composed of high-index materials on low-index backgrounds.[24–32] This geometry inherently breaks vertical symmetry and introduces radiative leakage through the substrate, degrading Q-factors and field confinement.[33–35] To suppress these losses, index-matching layers or polymer coatings are often applied,[36,37] but they reduce surface accessibility and limit the near-field coupling essential for atomic-scale integration.

Nanomembranes,[18–20,38] as thin and freestanding structures, exhibit excellent optical transparency, strong integration compatibility, and the ability to sustain high-Q resonances without substrate-induced losses. These attributes make them an ideal platform for overcoming the intrinsic limitations of substrate-supported BIC photonic structures. By entirely removing the substrate, these membrane structures restore out-of-plane symmetry and eliminate radiative leakage pathways, thereby preserving the non-radiative nature of BICs and enabling strong field confinement.[39–47] This suspended configuration maximizes light-matter interactions by supporting localized surface fields. More importantly, efficient optical interaction at the atomic scale requires photonic structures with thicknesses comparable to that of atomic-layer materials. Freestanding nanomembranes uniquely fulfill this requirement by enabling ultrathin architectures, ensuring dimensional compatibility and enhanced coupling efficiency.

In this work, we present an ultrathin and freestanding photonic nanomembrane that enables control of light coupling via atomic layer dielectric tuning with Å-level thickness precision (**Figure** 1a), and further demonstrate its capacity for 2D material integration by facilitating light–matter interactions in transition metal dichalcogenide (TMD) monolayers. By combining extreme thinness with substrate-free operation, it provides a versatile platform for integrating atomic-layer materials into advanced nanophotonic and quantum systems. Notably, sustaining high-Q resonances within such an ultrathin configuration represents a key advance for nanophotonic device performance. To understand the underlying mechanisms, we systematically investigate the influence of nanomembrane thickness on optical resonance behavior and identify substrate-induced field leakage as a primary factor degrading Q-factors.

Freestanding photonic nanomembranes, by eliminating this loss pathway, maintain exceptionally high Q-values across a wide spectral range, with Q = 9632 at 163 nm, Q = 3074 at 42 nm, and Q = 1480 at 29 nm. These results demonstrate that resonance quality can be preserved even when the membrane thickness is reduced into the deep subwavelength regime.[29] These findings underscore the critical role of freestanding architectures in overcoming optical losses and achieving robust high-Q resonances at extreme thicknesses. Beyond sustaining high-Q resonances, the ultrathin nanomembrane enables deterministic control of light coupling at the atomic scale. High-resolution spatial mapping of photoluminescence (PL) spectroscopy confirms that a single ALD cycle of silicon nitride (SiN, the dielectric materials same as the nanomembrane), corresponding to 0.65 Å in thickness, induces a 0.09 nm redshift (Figure 1d and 1e), and further reveals excellent spatial uniformity across the membrane. Furthermore, we extend the platform beyond dielectric tuning by integrating a $WS_2$ monolayer with the photonic nanomembrane. This integration results in strong field localization within the monolayer and a significant enhancement of emission, showing that ultrathin nanomembranes provide an effective route for controlling and enhancing light–matter interactions with 2D materials. This work establishes a versatile photonic platform for atomic-scale light-matter control, enabling strong field confinement, Å-level resonance tuning, and direct access to atomic-layer dielectrics and 2D materials for advanced nanophotonic and quantum applications.

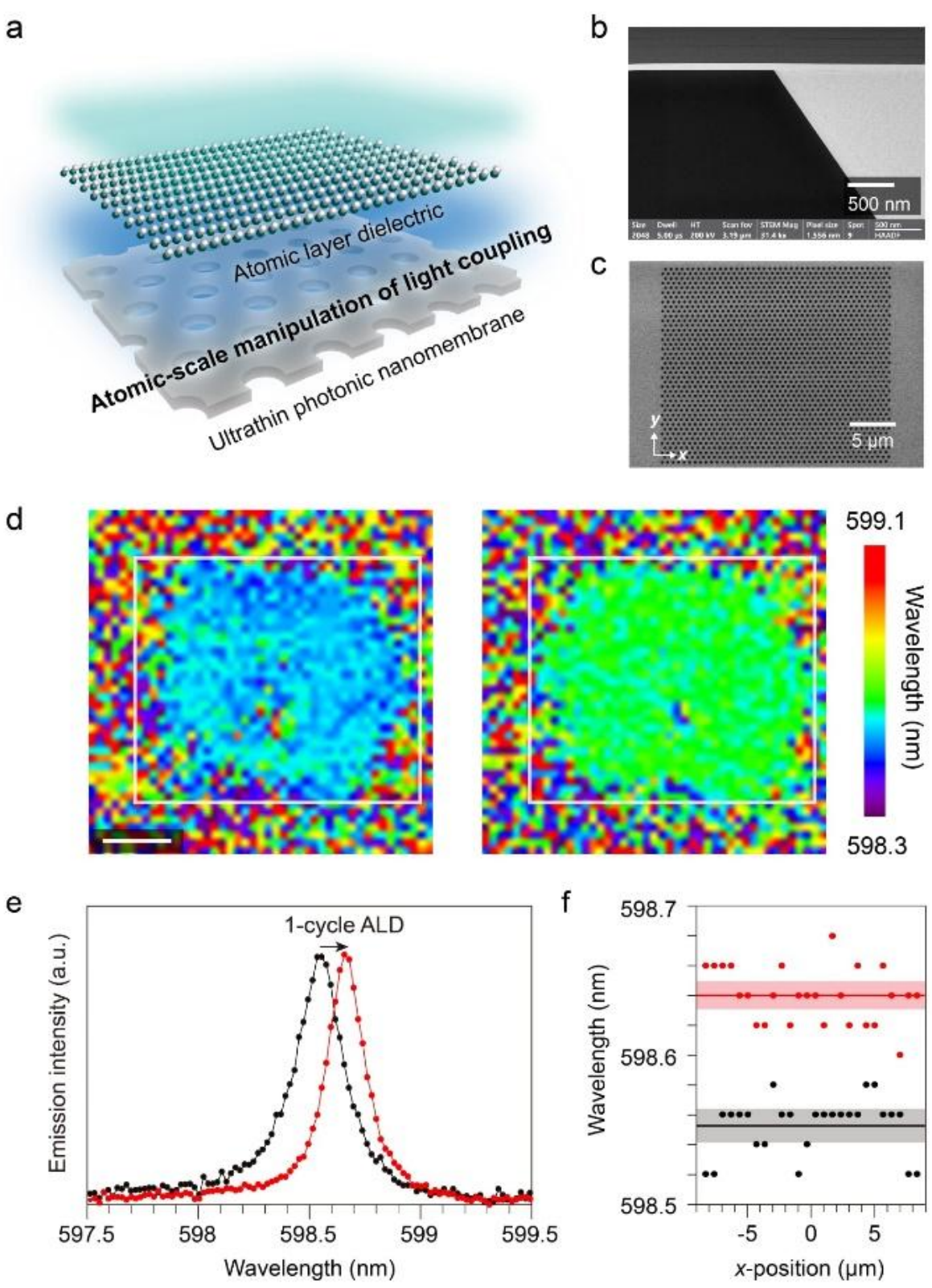

**Figure 1.** a) Schematic illustration of the ultrathin freestanding photonic nanomembrane designed for atomic-scale manipulation of light coupling and ultrasensitive optical characterization. Tailoring of mode resonances at the nanomembrane surface is realized through Å-level controlled dielectric deposition leveraging quasi-bound states in the continuum (quasi-BICs). b) Cross-sectional scanning transmission electron microscopy (STEM) image and c) top-view scanning electron microscopy (SEM) image depicting the structural morphology of the photonic nanomembrane with a Si-supported boundary. d) High-resolution spectral mapping and e) PL spectra demonstrating emission wavelength variations induced by a single ALD cycle of a dielectric layer. The white box marks the position of photonic nanomembrane. The scale bar represents 6 μm. f) Spatially resolved spectral shift profile along the *y*-direction, extracted from the emission spot, illustrating high uniformity and deterministic resonance tuning.

## 2. Design and Optical Characteristics of the Photonic Nanomembrane

The proposed SiN hole-array photonic nanomembrane is composed of a triangular lattice of airholes with lattice period $P$, diameter $D$ in a SiN nanomembrane of thickness $T$ (**Figure 2**a), where the coordinate system used throughout the manuscript is also indicated. Figure 2b presents the simulated angle-dependent reflectance spectra for the structure with $P$ = 600 nm, $D$ = 300 nm, and $T$ = 150 nm under *x*-polarized illumination. Numerical simulations were performed using rigorous coupled-wave analysis (RCWA) to characterize the structure. The incident light is tilted along the *y*-direction ($\theta_y$). The incident angle $\theta_y$ is kept between 0 and 5° in order to focus on the symmetry-protected BICs that are expected to arise at the Γ point. At normal incidence, the optical spectrum reveals modes exhibiting the theoretically infinite Q-factor characteristic of BICs, allowing for a clear distinction between guided modes (GMs) and BICs (**Figure S1**). However, in practical implementations, perfect BIC conditions are rarely achieved, resulting in quasi-BICs—modes that remain highly confined but still couple weakly to the radiative continuum. Despite this residual coupling, quasi-BICs retain exceptionally high Q-factors, making them particularly advantageous for applications that require strong light confinement while maintaining tunable interactions with the surrounding environment. Our analysis identifies two GMs and four BICs at normal incidence. Due to the nature of the triangular lattice, the observable optical modes differ depending on whether the incident angle is along the *x*- or *y*-direction. Additional simulations of the reflectance spectrum with light incident along the *x*-direction (**Figure S2**) reveal three more BICs ($BIC_{TM1}$, $BIC_{TM2}$, and $BIC_{TE2}$).

The distributions of electric energy densities $E_{den}$ for the resonances at 2° incident angle for GMs and quasi-BICs are presented in Figure 2c. Analysis of the electric field component distributions ($|E_x|$, $|E_y|$, $|E_z|$) enables the classification of transverse-electric (TE) and transverse-magnetic (TM) modes (**Figure S3**). The TM modes predominantly exhibit out-of-plane resonance characteristics, while TE modes demonstrate in-plane resonance behavior. Seven quasi-BICs were observed: three TM modes in the shorter wavelength range (~550–670 nm) and four TE modes in the longer wavelength range (~710–810 nm). The longer resonance wavelengths of the TE modes result from their stronger confinement within high-index regions, leading to a higher effective refractive index. Compared to GMs, which typically experience radiative losses and thus have finite Q-factors, quasi-BICs remain largely decoupled from free-space radiation near normal incidence. This weak radiative coupling enables them to sustain ultra-high Q-factors and significantly enhanced field intensities. Notably, $BIC_{TE4}$ exhibits an electric energy density approaching the order of $10^6$, whereas $GM_{TE}$ only reaches approximately $10^1$. Among TE quasi-BICs, $BIC_{TE4}$ is the only mode that exhibits resonance characteristics along both the primary lattice directions, while $BIC_{TE3}$ and $BIC_{TE2}$ (Figure S2) show electric energy densities on the order of $10^4$ and $10^3$, predominantly confined along the *x*- and *y*-directions, respectively.

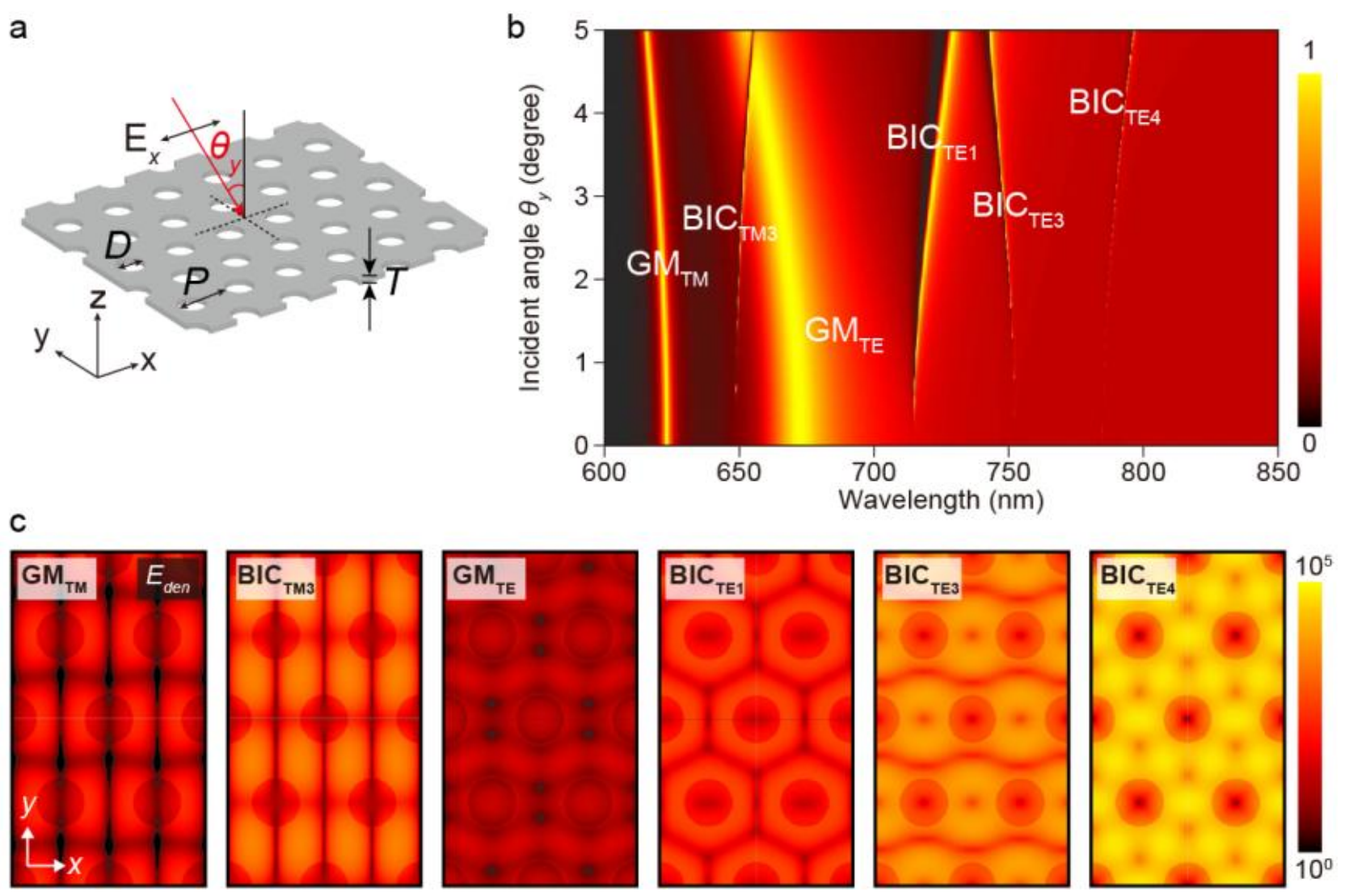

**Figure 2.** a) Schematic representation of the SiN photonic nanomembrane, comprising triangular-lattice air-hole array defined by lattice period $P$ = 600 nm, hole diameter $D$ = 300 nm, and nanomembrane thickness $T$ = 150 nm. b) Simulated angle-dependent reflection spectra under *x*-polarized illumination as the incident angle varies along the *y*-direction, revealing distinct optical resonances including GMs and quasi-BICs. c) Electric energy density ($E_{den}$)

distributions of the corresponding resonant modes identified in panel b), with field intensities presented on a logarithmic scale, highlight mode confinement and near-field enhancement characteristics.

## 3. Thickness-Dependent Mode Behaviors in Ultrathin Photonic Nanomembranes

To realize an ultrathin photonic nanomembrane, we next investigated the influence of nanomembrane thickness variation and the effect of substrate presence on the optical modes. Our analysis focused on how changes in the photonic slab thickness impact mode characteristics, including Q-factor, field enhancement, and field distribution. **Figure 3**a and e present simulated reflection spectra for two configurations: a SiN photonic slab on a $SiO_2$ substrate and a freestanding SiN photonic nanomembrane, respectively. Both configurations were analyzed across varying thicknesses under *x*-polarized light incidence at a 2° angle along the *y*-direction. In both configurations, all modes exhibited a blueshift with decreasing thickness, attributable to reduced effective optical path length. For the $SiO_2$-substrate configuration (Figure 3a), all resonances undergo attenuation, and the reflection peaks progressively lose contrast as the photonic slab thickness decreases. TM modes exhibit an even earlier suppression, with the $BIC_{TM3}$ becoming negligibly weak in the simulated response for thicknesses below 150 nm. The Q-factor variations with thickness for GMs and quasi-BICs are presented in Figure 3b. While the $BIC_{TE4}$ maintained a high Q-factor (~$10^4$), its resonance diminished to an undetectable level in the simulated response at approximately 100 nm thickness. The $GM_{TM}$ exhibited a relatively low Q-factor (~$10^2$). The extinction of both GMs and quasi-BICs at reduced thicknesses was attributed to strong field leakage into substrate radiation channels, preventing sustained resonance. In contrast, the freestanding nanomembrane configuration (Figure 3e) preserved notably high Q-factors and strong resonance even at thicknesses below 100 nm.

Unlike the substrate-based configuration, where the reflection intensity decreases with reducing thickness due to field leakage into the substrate, the reflection peaks of both GMs and quasi-BICs in the nanomembrane configuration remain high, indicating strong optical mode confinement and resonance. Notably, all TE modes remain observable even when the nanomembrane thickness is reduced below 10 nm. In contrast, for the two TM modes ($GM_{TM}$ and $BIC_{TM3}$), reflection peaks become difficult to resolve below a thickness of approximately 75 nm. However, these modes exhibit progressively narrower linewidths and increasing Q-factors as the nanomembrane thickness decreases, while their resonance remains strong. The inability to resolve the peaks arises from simulation limitations, specifically the finite wavelength resolution, which fails to accurately capture modes with extremely narrow bandwidths. This effect is distinct from the mode extinction observed in the substrate-based configuration near the cutoff thickness, where field leakage dominates (Figure 3a). The nanomembrane configuration provides a more stable optical environment for both GMs and quasi-BICs under reduced thickness conditions, in contrast to the substrate-based structure, where substrate-induced radiation losses significantly affect mode confinement.

Additionally, the nanomembrane configuration exhibits superior spectral tunability compared to the substrate-based structure, as evidenced by the more pronounced blueshift of resonances. Specifically, when the thickness is reduced from 200 nm to 100 nm, the $BIC_{TE4}$ exhibits a blueshift of 112 nm in the nanomembrane configuration, compared to 72 nm in the substrate configuration. Further reducing the thickness to 25 nm results in a total blueshift of approximately 270 nm from its original position at 200 nm thickness, demonstrating the significant spectral tuning capability of the nanomembrane structure. This superior spectral tunability primarily arises from two factors: the higher refractive index contrast between the photonic nanomembrane and air, and the absence of substrate coupling, which allows the mode to undergo greater changes in its dispersion characteristics. Figure 3f illustrates the simulated Q-factor variations with thickness for the nanomembrane configuration. For TM modes, both $GM_{TM}$ and $BIC_{TM3}$ exhibited an increase in Q-factor as the thickness decreased. Since the electric field along the *z*-direction dominates the resonance, TM-polarized modes exhibit greater sensitivity to thickness variations compared to TE modes. Furthermore, in a freestanding nanomembrane configuration, which eliminates substrate-induced radiation losses, TM modes achieve significantly higher Q-factors compared to their substrate-supported counterparts, especially for the thinner (< 100 nm) structure. This behavior corresponds to the stronger enhancement of the electromagnetic field at the photonic nanomembrane-air interface, rather than being confined within the photonic nanomembrane. Notably, the highest Q-factor observed for TM-quasi-BICs in the nanomembrane configuration reached $10^5$, two orders of magnitude higher than that of substrate-supported structures. Among the TE modes, the $BIC_{TE4}$ exhibited relatively high stability in its Q-factor, with it decreasing by only one order of magnitude as the thickness was reduced to 25 nm from 200 nm, while still maintaining a high value on the order of $10^3$. This higher Q-factor arises from its efficient resonance characteristics and high field enhancement as discussed in Figure 2b and 2c.

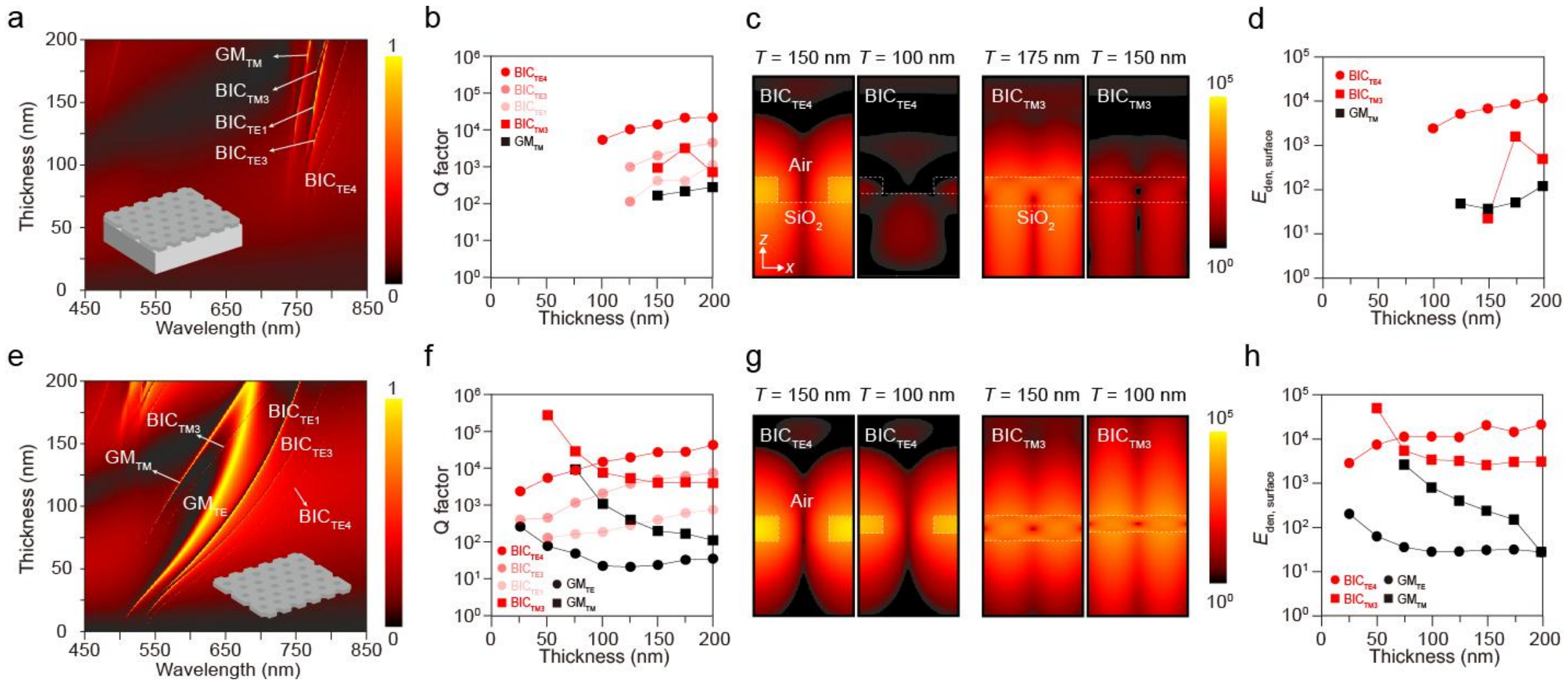

**Figure 3.** Thickness-dependent resonance and field confinement in substrate-supported and freestanding photonic nanomembranes. Simulated reflection spectra as a function of structural thickness for a) SiN hole-array photonic slabs supported on a $SiO_2$ substrate, and e) freestanding photonic nanomembranes. The incident light is *x*-polarized with an incident angle of 2° along the *y*-direction. Corresponding simulated Q-factor variations for GMs and quasi-BICs in b) substrate-supported photonic slabs and f) photonic nanomembranes. Electric energy density $E_{den}$ distributions in the *xz*-plane for the $BIC_{TE4}$ and $BIC_{TM3}$ in c) the substrate-supported photonic slabs and g) photonic nanomembranes at selected thicknesses. The *xz*-plane distributions are presented at the center and a position $y = 0.257$ µm offset from the center for the $BIC_{TE4}$ and $BIC_{TM3}$, respectively, corresponding to their maximum field intensities. Maximum $E_{den}$ values at the top membrane surface (*xy*-plane) of d) the substrate-supported photonic slabs and h) the photonic nanomembranes, highlighting significantly enhanced field confinement achievable in the ultrathin freestanding configuration.

## 4. Field Confinement in Ultrathin Photonic Nanomembranes

The field confinement and enhancement characteristics of quasi-BICs and GMs were further investigated as a function of the slab and nanomembrane thicknesses. Figure 3c and d focus on the photonic slab atop a $SiO_2$ substrate, while Figure 3g and h examine the nanomembrane case. Figure 3c displays the electric energy density distribution in the *xz*-plane of two modes: the $BIC_{TE4}$ at the thickness of 150 nm and 100 nm, and the $BIC_{TM3}$ at the thickness of 175 nm and 150 nm. As the thickness approaches 100 nm for the $BIC_{TE4}$ and 150 nm for $BIC_{TM3}$, significant field leakage into the substrate weakens field confinement, making it difficult to sustain the modes. Figure 3d illustrates the maximum electric energy density $E_{\text{den, surface}}$ at the top surface of the slab. Consistent with previously discussed Q-factor trends, quasi-BICs demonstrate stronger surface fields compared to GMs. The $BIC_{TE4}$ field intensity decreases with reduced thickness due to diminished confinement from reduced material volume. $BIC_{TM3}$ experiences even greater field leakage into the substrate due to the vertical resonance of the TM mode, resulting in weak field enhancement and barely sustaining at a relatively large thickness of 150 nm.

Figure 3g illustrates the electric energy density distributions in the *xz*-plane of the $BIC_{TM3}$ and $BIC_{TE4}$ for thicknesses of 150 nm and 100 nm for the nanomembrane case. Compared to the substrate-based configuration in Figure 3c, both modes retain strong field enhancement near the surface as the nanomembrane thickness decreases, benefiting from the absence of substrate-induced field leakage. The $E_{\text{den, surface}}$ for the photonic nanomembrane is plotted in Figure 3h. Due to their significantly higher Q-factors and lower radiative losses compared to GMs, the quasi-BICs exhibit higher surface field enhancements compared to GMs. The $BIC_{TE4}$ benefits from superior out-of-plane confinement and is consequently less sensitive to thickness variations; its surface field enhancement remains on the order of $10^4$ when thickness decreases from 200 nm to 75 nm, only declining to $10^3$ when further reduced to 25 nm. For the $BIC_{TM3}$, stronger surface field enhancement occurs with reduced thickness because its field is predominantly localized at the photonic nanomembrane-air interface rather than being strongly confined within the photonic nanomembrane. The surface field enhancement reaches a high value of approximately $10^5$ at a thickness of 50 nm. As previously noted, obtaining simulation results for structures thinner than this becomes increasingly challenging due to spectral resolution limitations in the simulation.

In summary, the proposed photonic nanomembrane maintains strong surface fields even with very small thicknesses. At 50 nm thickness, the $BIC_{TE4}$ and $BIC_{TM3}$ maintain impressive field enhancements of approximately $10^4$ and $10^5$, respectively. Notably, the $BIC_{TE4}$ demonstrates remarkable stability, decreasing by less than one order of magnitude when the thickness is reduced from 200 nm to 25 nm. In contrast, the presence of a substrate significantly disrupts and attenuates these modes due to field leakage, causing the optical mode resonance to become too weak to observe in thinner structures (< 100 nm). Furthermore, quasi-BICs consistently outperform GMs by approximately two orders of magnitude in surface field enhancement, underscoring their exceptional potential for applications in sensing applications and light-matter interactions within photonic structures and atomically thin 2D materials.

**5. High-Q Mode Resonances Tailoring in Ultrathin Photonic Nanomembranes**

To experimentally validate the theoretical predictions of the optical properties, optical characterization was first performed on a 200 nm-thick SiN photonic nanomembrane, serving as an initial demonstration before exploring thinner structures. The reflectance spectra were measured using a supercontinuum laser source (ROCK 480, LEUKOS, France). The light was focused using a 20× objective lens (NA = 0.25). The reflected light was collected and analyzed using a spectrometer (Kymera 328i Spectrograph, Oxford Instruments Group, UK) to obtain spectral measurements. Comprehensive details of the optical measurements are provided in the Methods section. **Figure 4**a shows the angle-resolved reflectance spectra measured along the *y*-direction. At normal incidence ($\theta_y = 0°$), symmetry-protected BICs remain uncoupled from far-field radiation due to their symmetric mode profiles. The spectrum at normal incidence exhibits two main features, with a narrow peak at 661.9 nm corresponding to the $GM_{TM}$ and a broader peak at 689.5 nm associated with the $GM_{TE}$. As the incident angle increases, quasi-BICs gradually couple to the far field, and at θ = 1.52° two modes ($BIC_{TE1}$ and $BIC_{TE3}$) become clearly observable. The measured reflection spectra of the SiN photonic nanomembrane exhibit consistency with simulated reflectance spectrum at near-normal incidence, confirming the presence of GMs and quasi-BICs as predicted by the simulations.

Because the SiN membrane used in this work exhibits defect-related states, a broad-band PL emission is observed under optical pumping (**Figure S4**), allowing the investigated GMs and quasi-BICs to couple to this PL band and exhibit high-Q resonances. This approach enables the probing of high-Q quasi-BICs through a PL emission peak, which improves the signal-to-noise ratio by reducing contributions from reflection and scattering. Such a method is advantageous for demonstrating light coupling control at the atomic-layer scale. Figure 4b and c present the emission spectra of the SiN photonic nanomembrane. The PL spectra of the SiN photonic nanomembranes were obtained utilizing a confocal Raman/PL microscope system (alpha300 R, WITec, Germany). The samples were excited with a CW laser at 488 nm. The excitation light was focused by objective lenses with a magnification of 10× (NA = 0.25). The excited emission was collected by the same objective lenses. The emission intensity distributions were obtained using a motorized *x*-*y*-sample scanning stage for confocal emission imaging. The nanomembrane was excited using a 488 nm continuous wave laser, and the resulting broadband PL emission was observed at approximately 500–800 nm (**Figure S4**). This emission is associated with defect-related states in SiN. The PL spectra of the SiN photonic nanomembrane exhibit high consistency with reflectance measurements at near-normal incidence, further confirming the presence of GMs and quasi-BICs as predicted by the theoretical analysis. For the GMs, the $GM_{TM}$ manifests as a well-defined peak, while the $GM_{TE}$ overlaps with adjacent quasi-BICs. Although this overlap complicates the direct identification of the $GM_{TE}$, its presence is confirmed through correlation with the reflectance measurements in Figure 4a. The three TM-quasi-BICs exhibit distinct spectral characteristics, with $BIC_{TM3}$ displaying the narrowest emission peak. This observation aligns well with the simulation results. However, due to the overlap of emission peaks, accurately resolving its Q-factor remains challenging. For the TE-quasi-BICs, although the spectral position of $BIC_{TE1}$ is shifted from the SiN PL band, it exhibits a high emission intensity comparable to that of the $GM_{TM}$, which resonates near 650 nm. The significant enhancement of emission (up to ~100-fold compared to a plain nanomembrane, as shown in **Figure S5**) demonstrates the potential of these structures for photonic applications in surface-emitting lasing, nonlinear optics, and quantum light generation. Notably, while $BIC_{TE4}$ exhibits relatively weak emission due to its wavelength being near the edge of the SiN PL band, it demonstrates an exceptionally high Q-factor of approximately 8065, as shown in Figure 4c. This measurement was obtained using a spectrometer grating with 1800 lines/mm, which provides significantly higher resolution compared to the 150 lines/mm grating used in Figure 4b, where the $BIC_{TE4}$ is difficult to resolve.

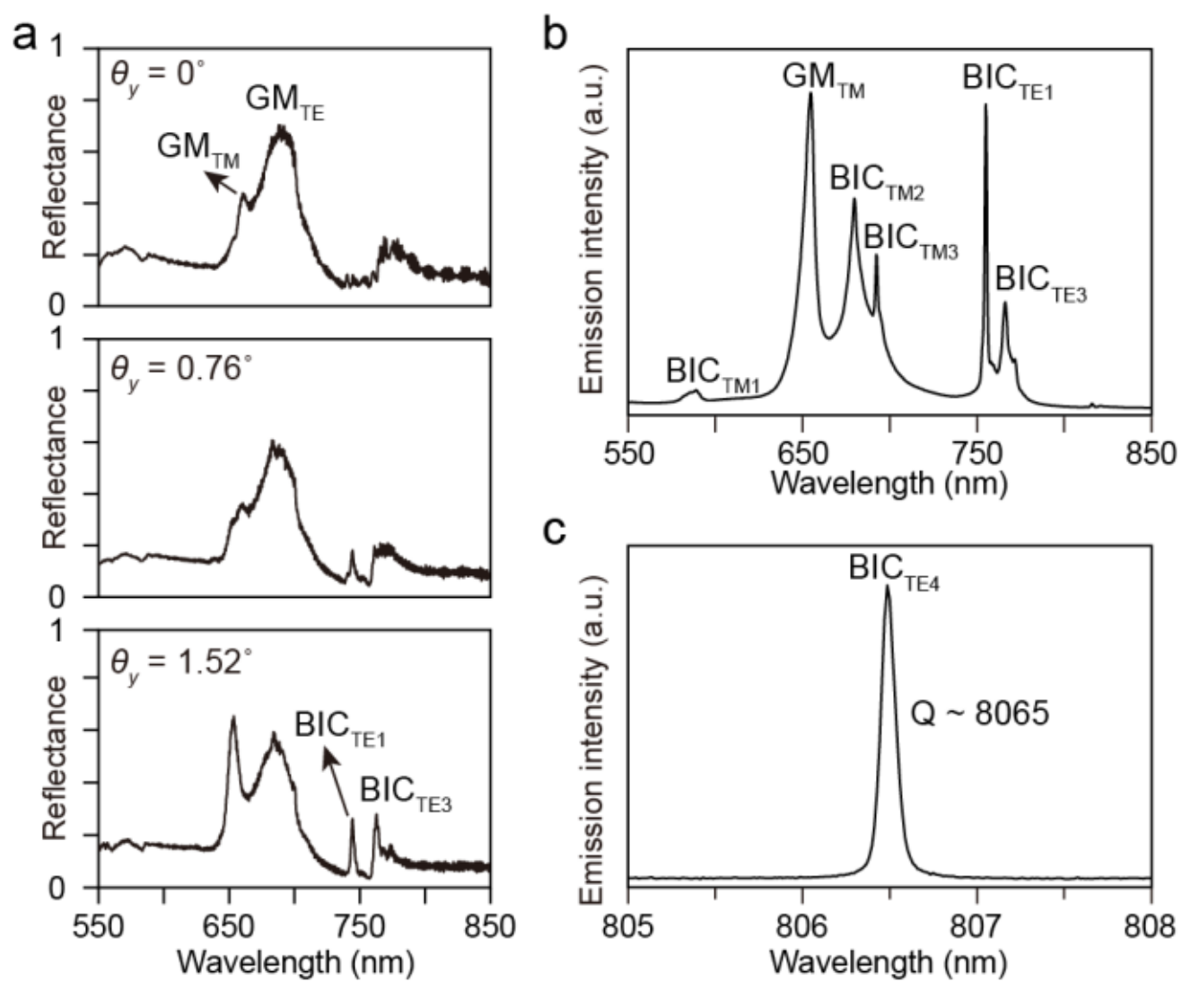

**Figure 4.** a) Reflectance spectra of the photonic nanomembrane with a thickness of 200 nm under varying incident angles. PL emission spectra from the SiN photonic nanomembrane, presented for b) the full spectral region of interest at lower spectral resolution and c) the $BIC_{TE4}$ region at higher spectral resolution, clearly resolving its high-Q resonance characteristics.

To extend the investigation to ultrathin nanomembranes, photonic nanomembranes with controlled thicknesses were developed. Building upon the initial study of a 200 nm-thick nanomembrane, the nanomembrane thickness was gradually reduced using a low-damage reactive ion etching (RIE) process with highly precise thickness modulation. This approach ensures exceptional structural preservation while enabling systematic exploration of optical properties in the ultrathin regime. The PL spectra of photonic nanomembranes with different thicknesses, measured using spectrometer gratings with 150 lines/mm and 1800 lines/mm, are presented in **Figure 5**a and 5b, respectively. The experimentally measured resonance wavelengths show strong agreement with our simulations, as demonstrated in Figure 5c and 5d. All modes exhibit the expected blueshift toward shorter wavelengths as the thickness decreases. In Figure 5b, the $BIC_{TE4}$ manifests as a remarkably sharp peak with an exceptional Q-factor of 9632 at a thickness of 163 nm. The Q-factor gradually decreases with reducing thickness but remains impressively high at 2037 and 1480 for thicknesses of 36 nm and 29 nm, respectively (**Figure S6**). The $BIC_{TE4}$ shift from 806.49 nm to 590.87 nm is demonstrated experimentally, spanning a substantial portion of the visible spectrum, exhibiting the great tunability of resonances by controlling the thickness. For TM modes, as the thickness decreases, their wavelengths shift further from the SiN PL emission band, making their observation more challenging in ultrathin structures. The $BIC_{TM3}$ remains observable only at a 163 nm-thick photonic nanomembrane, exhibiting a Q-factor of 3809 (**Figure S7**). Figure 5e and 5f present the simulated and measured Q-factors as a function of thickness, respectively. The TE quasi-BICs exhibit consistently higher Q-factors than TE GMs across all thicknesses, with $BIC_{TE4}$ displaying a Q-factor more than two orders of magnitude higher than $GM_{TE}$, as confirmed by both experimental and simulated results. Notably, the experimental results demonstrate that all $BIC_{TE4}$ maintain high Q-factors within the same order of magnitude as the thickness decreases, sustaining strong resonance even at 29 nm.

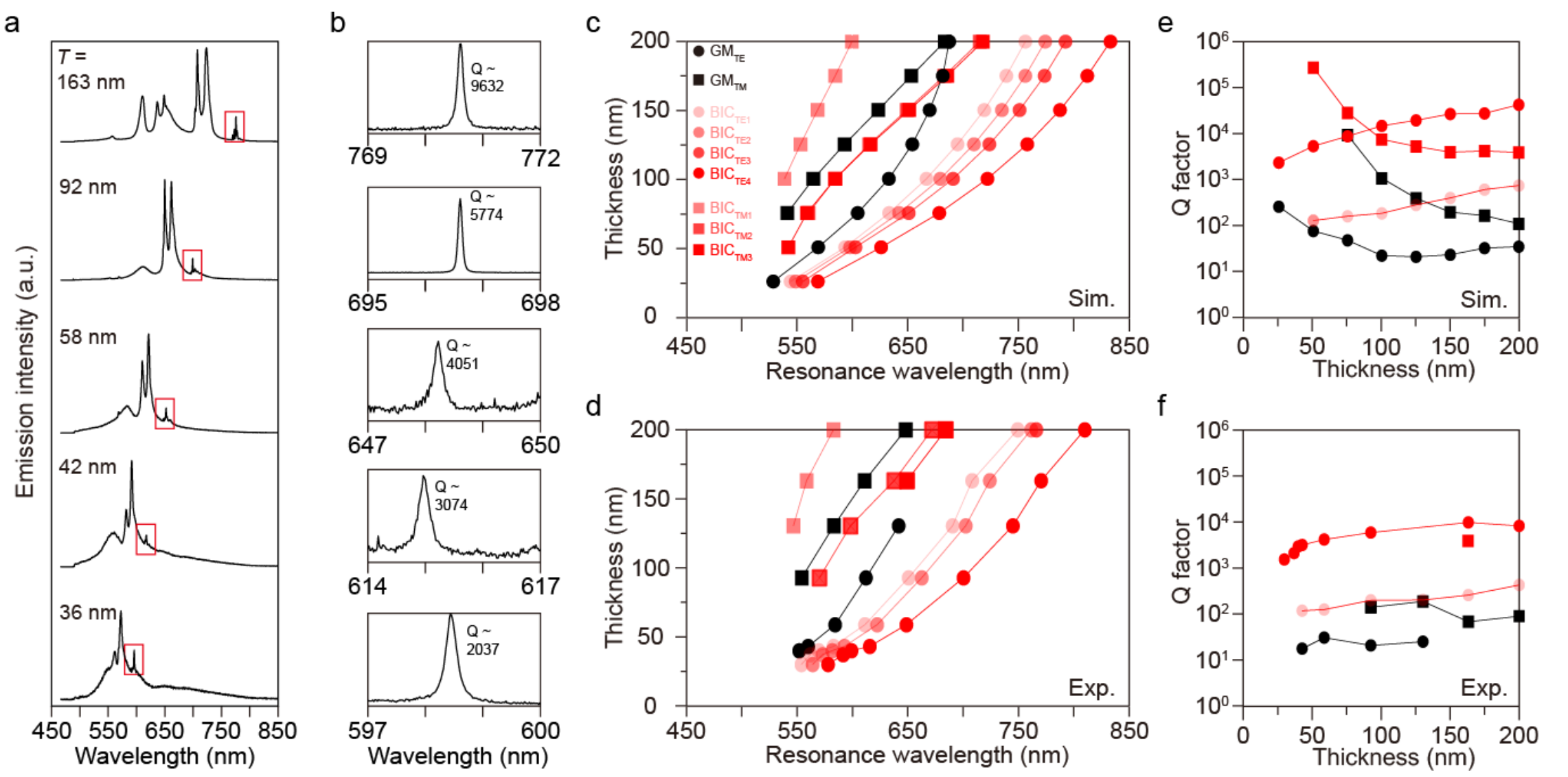

**Figure 5.** PL spectra of the photonic nanomembrane as a function of nanomembrane thickness, illustrating a) GMs and quasi-BICs, and b) specifically highlighting the high-Q $BIC_{TE4}$ region at higher spectral resolution. c) Simulated and d) experimentally measured resonance wavelengths of GMs and quasi-BICs as a function of nanomembrane thickness. Corresponding e) simulated and f) experimentally measured Q-factors for the GMs and quasi-BICs, demonstrating sustained high-Q resonances achievable in the ultrathin nanomembrane regime.

While numerous photonic structures exhibiting high-Q quasi-BICs have been reported, the majority of these demonstrations have been conducted in the near-infrared region, with relatively few studies achieving high Q-factors in the visible range. Table S1 presents a comparison of experimentally obtained Q-factors in nanophotonic structures operating in the visible range. Our freestanding structure demonstrates Q-factors of $10^3$ in the visible wavelength range while maintaining an exceptionally thin layer of just 29 nm. In this wavelength range, most reported approaches to achieving high Q-factors in photonic structures supporting quasi-BICs have relied on either thick suspended photonic nanomembranes[42,44] or substrate-based nanoarrays. [27–31,36] While the former requires substantial thickness to ensure sufficient field confinement, the latter typically suffers from reduced Q-factors due to additional leaky channels introduced by out-of-plane asymmetry. The presented freestanding photonic nanomembrane achieves and sustains high-Q resonances in the visible region at an ultrathin thickness, representing a significant advancement in miniaturization without compromising performance. Furthermore, the finite-size effect generally leads to Q-factor degradation,[47,48] with high-Q modes often observed in larger sample sizes. Despite these limitations, our structure maintains strong optical confinement even within a compact footprint of just 25 μm × 25 μm. Notably, most measurements of quasi-BICs reported in the visible spectral regime have been characterized through transmission or reflection measurements,[27,29,31,44] or via lasing signals under high-power excitation in conjunction with optical gain media.[28,30,31] In contrast, our approach demonstrates high-Q quasi-BICs through a spontaneous emission peak excited with low pump power. This method provides an improved signal-to-noise ratio and enhanced detection limits while minimizing the risk of sample damage associated with high-power excitation, making it well suited for applications that rely on high-Q resonances.

## 6. Atomic Layer Controlled Light Coupling in Ultrathin Photonic Nanomembranes

To achieve such atomic-level control over light coupling at the nanomembrane surface, atomically thin dielectrics are integrated via ALD, ensuring atomic-scale precision in thickness control and interface quality (**Figure 6**). ALD was performed on the nanomembrane using a plasma-enhanced ALD system (AD-230LP, SAMCO Inc., Japan), with SiN selected as the deposited material at a deposition rate of 0.65 Å per cycle. (see Method) Figure 6a presents the PL spectra of the 48 nm-thick nanomembrane before ALD deposition and after undergoing 8 ALD cycles, followed by an additional 2 cycles (totaling 10 cycles), corresponding to the $BIC_{TE4}$. This measurement was obtained using a spectrometer grating with 1800 lines/mm. The initial PL emission peak was observed at 596.94 nm. After 8 ALD cycles, the $BIC_{TE4}$ red-shifted to 598.01 nm, corresponding to a shift of 1.07 nm for an estimated deposited layer thickness of 5.2 Å. With an additional 2 cycles, the mode further red-shifted to 598.31 nm, exhibiting an incremental shift of 0.30 nm upon the deposition of an extra 1.3 Å. Figure 6b presents the spatial mapping of the PL emission peak wavelengths of the $BIC_{TE4}$ across the nanomembrane, while Figure 6c shows the corresponding peak wavelengths for each mapped pixel, extracted from the emission region marked with dashed boxes in Figure 6b. This measurement was obtained using a spectrometer grating with 600 lines/mm. An 8 × 8-pixel region (corresponding to 16 μm × 16 μm) exhibits high uniformity, confirming that the PL peak shifts induced by the ALD process are robust and consistent across the photonic nanomembrane. The emission peak variation induced by a single-cycle ALD is further demonstrated in Figure 1e. A red shift of 0.09 nm was confirmed for a single ALD cycle, attributed to the high Q-factor of 2992, corresponding to an estimated atomic-layer thickness of 0.65 Å (see Method). Furthermore, high-resolution spatial mapping of the emission peak wavelength exhibits excellent uniformity, verifying that sub-Å atomic-level control over light coupling at the ultrathin nanomembrane surface has been successfully achieved. The inset of Figure 6a presents the simulated and measured shift (Δλ) of the $BIC_{TE4}$ induced by the deposition of a dielectric layer on a 48 nm-thick nanomembrane surface. Numerical simulations predict (inset of Figure 6a) a redshift consistent with experimental results, confirming the strong agreement between theory and measurement. It is noted that $T_{ALD}$ represents the estimated ALD thickness used in the simulation.

In Figure 6d-g, we further analyze the near-field enhancement in both thin ($T$ = 48 nm) and thick ($T$ = 200 nm) photonic nanomembranes, with and without the deposition of a 2-cycle ALD layer (1.30 Å). The normalized field enhancement distributions along the $z$-direction at the midpoint between two air holes in the $x$-direction are compared for both cases. The 48 nm-thick nanomembrane exhibits a significantly larger variation in surface field enhancement compared to the 200 nm-thick structure. Additionally, for the ultrathin nanomembrane, the surface field density exhibits a greater relative change upon deposition, highlighting its enhanced sensitivity to atomic-scale modifications. This amplified field response in the ultrathin nanomembrane is primarily attributed to the larger ratio of the deposited dielectric layer thickness relative to the overall nanomembrane thickness. This demonstrates the advantage of ultrathin photonic nanomembranes for achieving enhanced light coupling with atomic-layer dielectrics and monolayer 2D materials, where precise control over surface optical interactions is critical.

It is noted that atomic-layer dielectrics pose challenges for atomic-level light coupling due to their low refractive index and minimal extinction coefficient, leading to weak optical contrast. In contrast, 2D semiconductors, such as TMD monolayers, owing to their high refractive indices and strong extinction coefficients, show pronounced resonance variations under atomic-scale modifications, as discussed in the introduction and demonstrated in **Figure 7**. Our demonstration highlights the capability of the ultrathin photonic nanomembrane system to resolve subtle optical variations within atomic-layer dielectrics at the nanomembrane surface. This ensures that the proposed ultrathin photonic nanomembrane can achieve significantly enhanced atomic-scale light coupling when integrated with other 2D materials or low-dimensional material systems, providing a robust and versatile platform for next-generation nanophotonic and quantum applications.

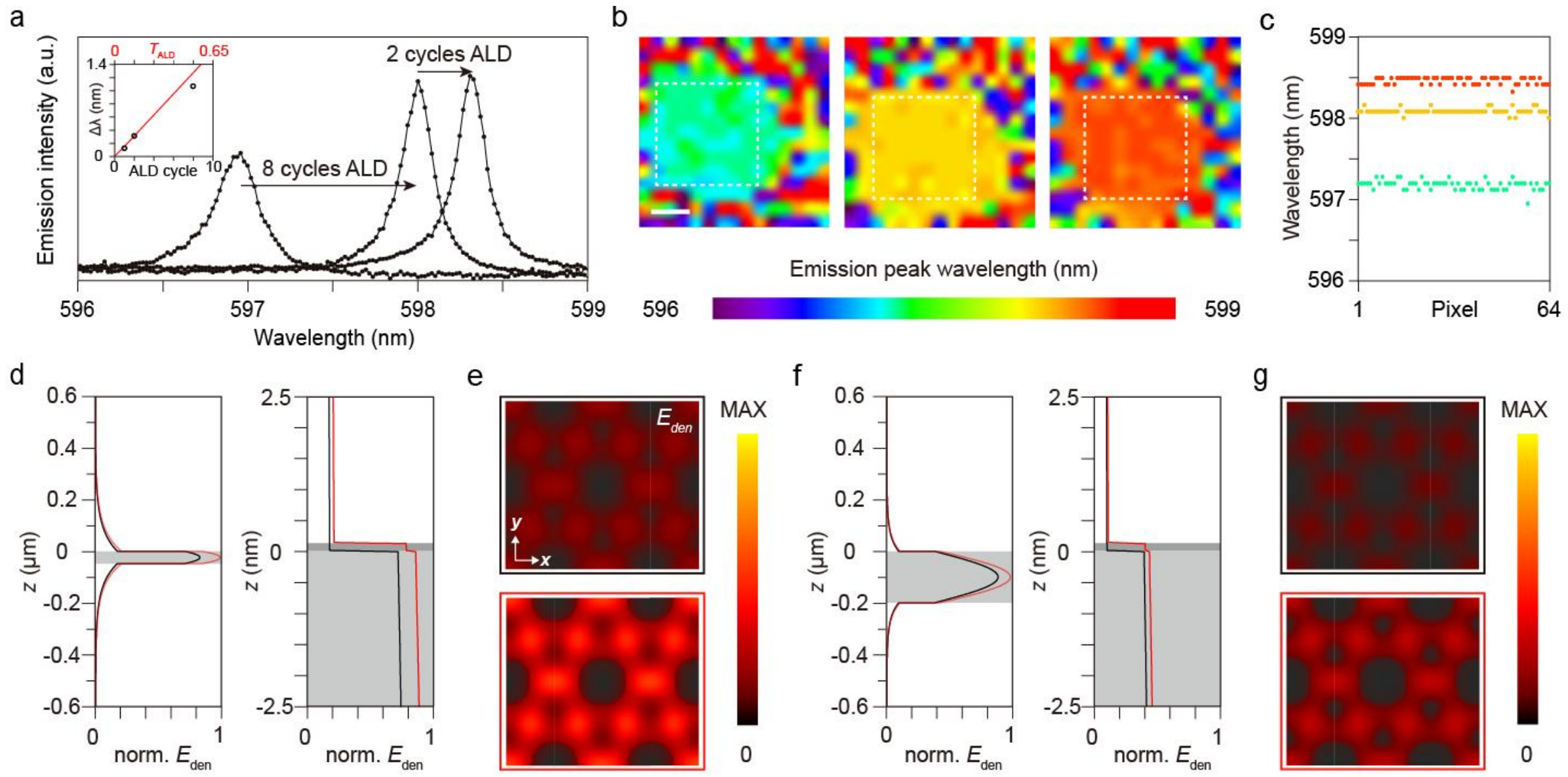

**Figure 6.** a) PL spectra demonstrating spectral shifts of the $BIC_{TE4}$ in the 48-nm-thick photonic nanomembrane induced by atomic-scale dielectric depositions through ALD. The inset compares simulated and experimentally measured $BIC_{TE4}$ emission shifts due to ALD, where $T_{ALD}$ represents the estimated ALD thickness used in the simulation. b) Spatially resolved mapping of the $BIC_{TE4}$ emission peak wavelengths before and after atomic-layer depositions, demonstrating uniform and deterministic wavelength shifts across the nanomembrane surface. The scale bar represents 6 µm. c) Emission peak wavelengths of the $BIC_{TE4}$ for each mapped pixel, extracted from the emission region marked with dashed boxes in b). Comparative analysis of field enhancement using normalized $E_{den}$ distributions along the *z*-direction for d) 48-nm and f) 200-nm thick photonic nanomembranes, before ALD deposition (black) and after a 1.30-Å dielectric layer deposition (red). The right panels provide detailed views of the nanomembrane surface facing the incident light. The *xy*-plane field distribution variations before and after ALD for e) 48-nm and g) 200-nm thick nanomembranes at the same *z*-position indicated in d) and f). The field distributions illustrate enhanced near-field interactions attainable in the ultrathin nanomembrane configuration.

## 7. Ultrathin photonic nanomembrane for enhanced light–matter interaction in TMD monolayers

Building on the atomic-level light-coupling control enabled by ALD-based dielectric tuning, the platform with integrated 2D material is subsequently investigated. Figure 7 illustrates the enhanced light-matter interaction achieved by integrating a 2D material ($WS_2$ monolayer) with the photonic nanomembrane. The details of the synthesis of the $WS_2$ monolayer are provided in the Methods section. Here, the resonance variations induced by atomic-layer SiN and by integration of a $WS_2$ monolayer on the photonic nanomembrane are first investigated and compared. Figure 7a presents the simulated spectra of a 58-nm-thick photonic nanomembrane, the same membrane with 8-cycle SiN ALD (0.52 nm thick), and the same membrane with integration of a $WS_2$ monolayer (0.6 nm thick). It should be noted that, due to the strong extinction coefficient of $WS_2$ near its excitonic transition,[49] reflection spectra do not clearly reveal the resonances; therefore, we instead simulated the absorptance spectrum of the $WS_2$-integrated photonic nanomembrane. The resonance wavelengths of the bare nanophotonic membrane are identified at 579.87 nm ($GM_{TE}$), 580.90 nm ($BIC_{TE1}$), 617.72 nm ($BIC_{TE3}$), and 642.52 nm ($BIC_{TE4}$). Deposition of 8 SiN-ALD cycles redshifts these resonances by 1.03, 1.22, 1.26, and 1.25 nm, respectively.

In contrast, integration of the $WS_2$ monolayer induces much larger redshifts of 5.99 nm for $GM_{TE}$, 12.10 nm for $BIC_{TE3}$, and 10.58 nm for $BIC_{TE4}$. Notably, the $BIC_{TE1}$ resonance exhibits mode splitting in the PL spectrum (**Figure S8**), a clear signature of strong coupling between the TMD exciton and the quasi-BICs.[33, 41, 44] This difference highlights that integration with a 2D material such as a $WS_2$ monolayer leads to more significant resonance variations. More importantly, the coupling between the quasi-BIC modes and the excitonic response of $WS_2$ gives rise to enhanced light–matter interaction and the onset of strong exciton–photon coupling, as evidenced by the observed mode splitting. Figure 7b shows the spatial distribution of the electric field energy density at the two polariton branches resulting from Rabi splitting of the exciton resonance coupled to $BIC_{TE1}$ (605.53 nm and 620.09 nm) and at $BIC_{TE3}$ (629.82 nm) . The field is strongly amplified and tightly confined within the $WS_2$ monolayer near the top surface instead of the photonic nanomembrane. This confirms the strong exciton–photon coupling arising from the large spatial overlap between the quasi-BIC resonance field and the $WS_2$ monolayer. The resulting enhanced light–matter interaction leads to a pronounced increase in emission. In Figure 7c, emission spectrum from a $WS_2$ monolayer placed on the photonic nanomembrane is substantially enhanced compared to that of the same monolayer placed on the unstructured nanomembrane area. Figure 7d quantifies the emission enhancement using the enhancement factor, defined as the ratio of PL from $WS_2$ on the photonic nanomembrane to that on unstructured membranes, which reaches a maximum of 113-fold at the coupled resonance wavelength. The observed strong enhancement suggests the upper and lower polariton branches resulting from exciton–BIC hybridization. These results demonstrate that ultrathin photonic nanomembranes provide an effective platform for enhancing light–matter interactions in the 2D material at the atomic scale. It is noted that through appropriate photonic geometry design, the ultrathin photonic nanomembrane can be integrated with a wide range of 2D materials, as discussed in the Supporting Information (**Figure S9**).

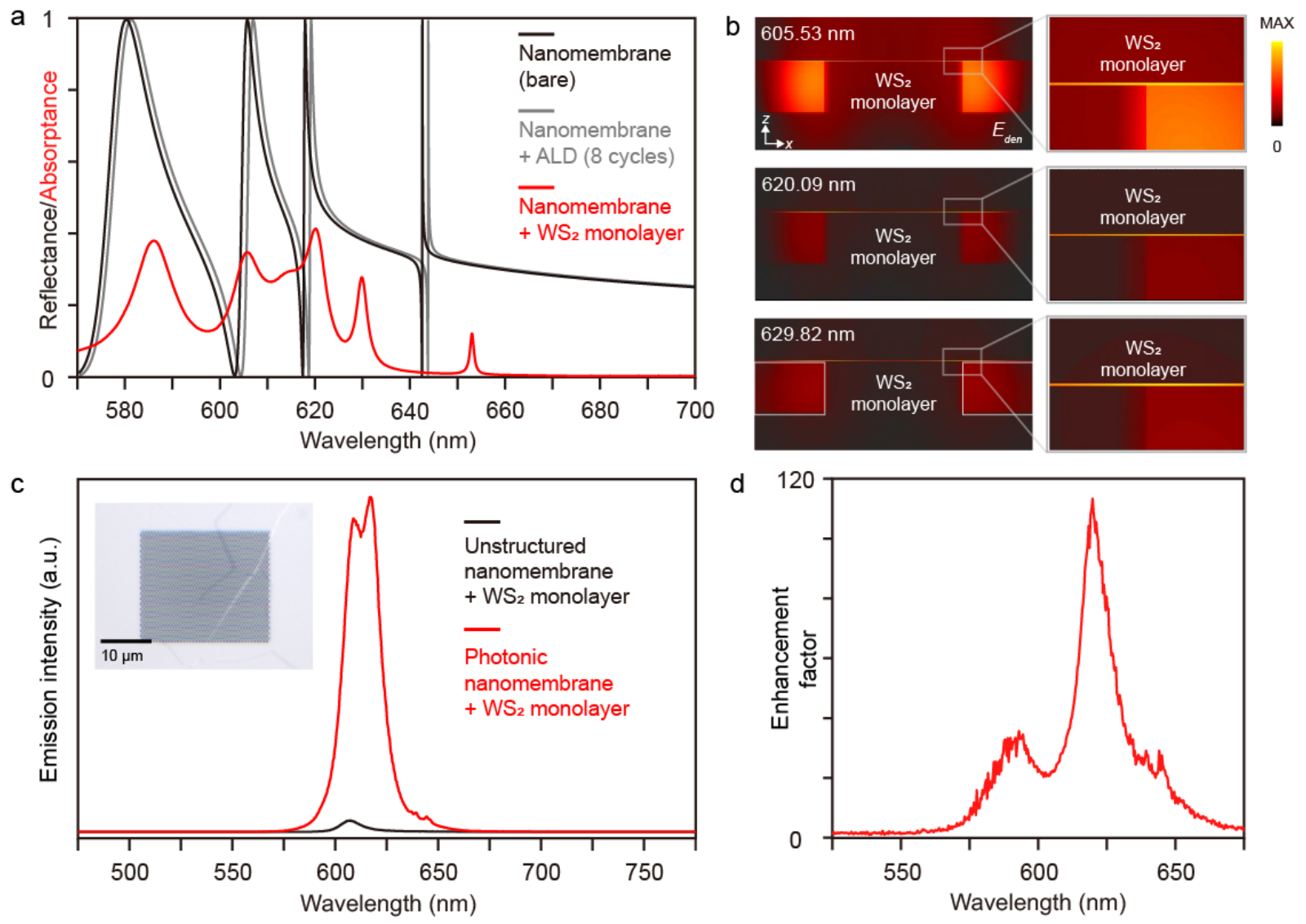

**Figure 7.** a) Simulated reflectance spectra of a 58-nm-thick photonic nanomembrane before and after 8 cycles of ALD of SiN, together with the simulated absorptance of the same membrane with a $WS_2$ monolayer on top. b) Spatial distribution of the electric field energy density in *xz*-plane for the quasi-BICs with a $WS_2$ monolayer integrated onto the photonic nanomembrane. c) PL spectra of a $WS_2$ monolayer on an unstructured nanomembrane compared with that on the photonic nanomembrane. Inset: Optical microscope image of the photonic nanomembrane with the transferred $WS_2$ monolayer. d) Enhancement factor, defined as the PL intensity ratio between the $WS_2$ monolayer on the photonic nanomembrane and that on the unstructured membrane, together with the PL spectrum of the bare 58-nm-thick nanomembrane prior to $WS_2$ transfer.

## 8. Conclusion

This work demonstrates an ultrathin freestanding photonic nanomembrane that enables atomic-level control over light coupling between light incidence and the quasi-BIC modes of the photonic nanomembrane, achieving Å-scale modulation of high-Q optical resonances. Through a systematic investigation of both GMs and quasi-BICs, we reveal that these ultrathin nanomembranes sustain high-Q factors, maintaining 3074 at 42 nm and 1480 even at 29 nm. By eliminating substrate-induced losses, this design preserves strong optical confinement and near-field enhancement, overcoming the inherent limitations of conventional nanophotonic structures. Furthermore, the integration of atomically thin dielectrics via ALD confirms the sub-Å precision of light coupling and atomic-layer characterization, with a single ALD cycle (0.65

Å) inducing a 0.09 nm red shift. High-resolution spatial mapping verifies the uniformity of this shift, demonstrating robust and deterministic control over atomic-scale light-matter interactions. In contrast to thicker photonic structures, ultrathin nanomembranes exhibit pronounced sensitivity of their resonant modes and field distributions to surface modifications at the Å scale, making them promising platforms for coupling with atomic-layer dielectrics, 2D materials, and other low-dimensional quantum systems. The pronounced PL enhancement together with the exciton–polariton mode splitting further demonstrates the strong light–matter interaction achieved in the 2D material–photonic nanomembrane platform. By establishing a scalable and highly tunable photonic platform, this work not only advances nanophotonic technologies, including quantum photonics, nonlinear optics, and ultra-sensitive optical sensing but also provides a powerful approach for sub-Å precision material characterization, which is critical for the most advanced semiconductor technologies.

## 9. Methods

*Simulation Details*

The far-field reflectance spectra and dispersion diagram of the photonic nanomembrane, as well as the near-field electric and magnetic field distributions of the resonance modes, were computed using the rigorous coupled-wave analysis (DiffractMOD, RSoft Design Group, USA). All simulations were performed under periodic boundary conditions in the $x$- and $y$-axes and perfectly matched-layer conditions in the $z$-axis, with plane-wave light incidence and the incident angle along the $z$-axis. The electric field $\mathbf{E}$ is normalized by the electric field amplitude of the incident light. The electric energy density is defined as $U_E = \frac{1}{2} \int \mathrm{Re}[\varepsilon(\mathbf{r}')] |\mathbf{E}|^2 dV$, where $\mathbf{E}$ is the electric field, $\varepsilon$ is the spatially dependent permittivity, and $V$ is the volume of the simulation grid. The simulation results are normalized by the input power. The refractive index of SiN was modeled according to literature values.[50]

*Optical Measurement*

The reflection spectra of the SiN photonic nanomembranes were measured using a supercontinuum laser source (ROCK 480, LEUKOS, France). The SiN nanomembranes are provided by Ted-Pella Inc. and further nanofabricated for the ultrathin nanomembrane samples utilized in this work. The light was focused using a 10× objective lens. The reflected light was collected and analyzed using a spectrometer (Kymera 328i Spectrograph Oxford Instruments Group, UK) to obtain the spectral measurements. The PL spectra of the SiN photonic nanomembranes were obtained utilizing a confocal laser microscope system (alpha300 R,

WITec, Germany). The samples were excited with a CW laser at 488 nm. The excitation light was focused by objective lenses with a magnification of 10× (NA = 0.25). The excited emission was collected and analyzed with a spectrometer. The emission intensity distributions were obtained using a motorized *x*-*y*-sample scanning stage for confocal emission imaging.

*Characterization of Freestanding Nanomembrane*

The freestanding ultrathin SiN nanomembranes were first treated with carbon deposition, with a thickness of 450 nm on the surface and 225 nm on the backside, to enhance the mechanical stability of the nanomembrane and suppress charge buildup. A room-temperature curing epoxy resin was then prepared by mixing the base and curing agents in a 3:1 ratio, followed by thorough stirring and degassing before being injected into the backside of the nanomembrane to provide structural support for the freestanding film. The resin was cured at 60°C for 5 hours to ensure structural stability. Focused ion beam scanning electron microscopy (FIB-SEM) (Ethos NX5000, Hitachi High-Tech, Japan) was then used to identify the interface between the SiN nanomembrane and the Si support frame, enabling precise extraction of the nanomembrane for scanning transmission electron microscopy (STEM) (Talos F200X G2, Thermo Fisher Scientific, USA) cross-sectional analysis. The sample for the observation was carefully separated from the support frame and mounted onto a TEM grid. The sample was thinned to below 100 nm, after which the final film thickness was verified using TEM imaging. The SEM images of etched photonic membranes confirms that the thinning process does not compromise the fidelity of the original photonic nanomembrane design (**Figure S10**).

*Atomic Layer Deposition*

SiN thin films were deposited using an ALD process with bis(diethylamino)silane (BDEAS) as the silicon precursor and nitrogen plasma ($N_2$ plasma) as the reactant. The deposition was carried out at a substrate temperature of 350°C under a chamber pressure of 8.8 Pa. In this work, a small number of ALD cycles (1–8 cycles) were performed to achieve atomic-scale dielectric control. To estimate the atomic-layer thickness per ALD cycle, calibration data from 2400 ALD cycles was referenced. The total deposited film thickness of 145.76 nm yielded a growth per cycle (GPC) of 0.65 Å/cycle, consistent with previous ALD calibration studies. Each cycle consisted of precursor exposure, plasma activation, and purge steps, with a total duration of 25 seconds per cycle. The refractive index of the deposited SiN film was characterized to be 1.934.

*Preparation for $WS_2$ monolayer*

$WS_2$ films were synthesized by conventional chemical vapor deposition (CVD) in a one-zone horizontal 2-inch quartz tube furnace. Tungsten trioxide ($WO_3$, Sigma-Aldrich, 99.9%, 100 mg) was used as the tungsten precursor and placed at the center zone together with c-plane sapphire substrates. The center zone was heated to 960 °C and maintained for 15 min for film growth. Sulphur powders (S, Sigma-Aldrich, 99.99%, 3 g), which were pre-solidified before use, were placed upstream in a quartz boat and heated by an external heating belt to 145 °C. The sulphur heating was initiated 10 min prior to the central zone reaching the target growth temperature. Before growth, the furnace tube was evacuated to base pressure and then purged with a mixed carrier gas consisting of Ar (200 sccm) and $H_2$ (40 sccm) at a constant pressure of 20 torr. After growth, the system was naturally cooled to room temperature under the Ar/$H_2$ flow.

The transfer of $WS_2$ films was carried out using a PDMS-assisted wet etching approach. A home-fabricated polydimethylsiloxane (PDMS) stamp was laminated onto the $WS_2$ film surface by natural adhesion. The sapphire substrate was then etched away in KOH solution, releasing the $WS_2$ film onto the PDMS. The PDMS/$WS_2$ stack was subsequently brought into contact with the photonic nanomembrane and placed under vacuum for 1 h to enhance the adhesion between the $WS_2$ film and the photonic nanomembrane. Afterwards, mild heating at 60 °C for 10 min was applied to facilitate the detachment of $WS_2$ from the PDMS, leaving the film successfully transferred onto the photonic nanomembrane.

**Acknowledgements**

This work was supported by JSPS KAKENHI Grant Number JP23K26155, JP21H04660. A part of this work was supported by Advanced Research Infrastructure for Materials and Nanotechnology in Japan (ARIM) of the Ministry of Education, Culture, Sports, Science and Technology (MEXT) (Proposal Number JPMXP1225NM5090). Financial support by the Center of Atomic Initiative for New Materials (AI-Mat), National Taiwan University, from the Featured Areas Research Center Program within the framework of the Higher Education Sprout Project by the Ministry of Education in Taiwan (Grant Number 111L900801), is also acknowledged. The authors would like to extend our grateful appreciation to Dr. Takuro Nagai, Dr. Noriyuki Okada, Dr. Makoto Oishi, and Dr. Rika Mizuta from the Electron Microscopy Unit, Research Network and Facility Services Division, National Institute for Materials Science NIMS, for important technical support on TEM.

**Data Availability Statement**

The data that support the findings of this study are available from the corresponding author upon reasonable request.

**References**

[1] Loh, L., Wang, J., Grzeszczyk, M., Koperski, M., Eda, G., Towards quantum light-emitting devices based on van der Waals materials, 2024, *Nat. Rev. Electr. Eng.*, 1, 815–829, https://doi.org/10.1038/s44287-024-00108-8

[2] Song, S., Rahaman, M., Jariwala, D., Can 2D Semiconductors Be Game-Changers for Nanoelectronics and Photonics?, 2024, *ACS Nano*, 18, 10955, https://doi.org/10.1021/acsnano.3c12938

[3] Ahmed, T., Zha, J., Lin, K. K., Kuo, H. C., Tan, C., Lien, D. H., Bright and Efficient Light‐Emitting Devices Based on 2D Transition Metal Dichalcogenide, 2023, *Adv. Mater.*, 31, 2208054, https://doi.org/10.1002/adma.202208054

[4] Lee, Y. C., Chang, S. W., Chen, S. H., Chen, H. L., Optical Inspection of 2D Materials: From Mechanical Exfoliation to Wafer-Scale Growth and Beyond, 2022, *Adv. Sci.*, 9, 2102128, https://doi.org/10.1002/advs.202102128

[5] Roy, S., Zhang, X., Puthirath, A., Meiyazhagan, A., Bhattacharyya, S., Rahman, M., Babu, G., et al., Structure, Properties and Applications of Two-Dimensional Hexagonal Boron Nitride, 2021, *Adv. Mater.*, 33, 2101589, https://doi.org/10.1002/adma.202101589

[6] Bae, S. H., Kum, H., Kong, W., Kim, Y., Choi, C., Lee, B., Lin, P., et al., Integration of bulk materials with two-dimensional materials for physical coupling and applications, 2019, *Nat. Mater.*, 18, 550, https://doi.org/10.1038/s41563-019-0335-2

[7] Guo, Q., Qi, X. Z., Zhang, L., Gao, M., Hu, S., Zhou, W., Zang, W., et al., Ultrathin quantum light source with van der Waals NbOCl2 crystal, 2023, *Nature*, 613, 53, https://doi.org/10.1038/s41586-022-05393-7

[8] Lee, M., Hong, H., Yu, J., Mujid, F., Ye, A., Liang, C., Park, J., Wafer-scale δ waveguides for integrated two-dimensional photonics, 2023, *Science*, 381, 648, https://doi.org/10.1126/science.adi2322

[9] Li, Q., Song, J. H., Xu, F., van de Groep, J., Hong, J., Daus, A., Lee, Y. J., et al., A Purcell-enabled monolayer semiconductor free-space optical modulator, 2023, *Nat. Photonics*, 17, 897, https://doi.org/10.1038/s41566-023-01250-9

[10] van de Groep, J., Song, J. H., Celano, U., Li, Q., Kik, P. G., Brongersma, M. L., Exciton resonance tuning of an atomically thin lens, 2020, *Nat. Photonics*, 14, 426, https://doi.org/10.1038/s41566-020-0624-y

[11] Dorsey, K. J., Pearson, T. G., Esposito, E., Russell, S., Bircan, B., Han, Y., Miskin, M. Z., et al., Atomic Layer Deposition for Membranes, Metamaterials, and Mechanisms, 2019, *Adv. Mater.*, 31, 1901944, https://doi.org/10.1002/adma.201901944

[12] Lien, D. H., Uddin, Z., Yeh, M., Amani, M., Kim, H., Allen, J. W. A. III, Yablonovitch, E., Javey, A., Electrical suppression of all nonradiative recombination pathways in monolayer semiconductors, 2019, *Science*, 364, 468, https://doi.org/10.1126/science.aaw8053

[13] Li, Y., Zhang, J., Huang, D., Sun, H., Fan, F., Feng, J., Wang, Z., Ning, C. Z., Room-temperature continuous-wave lasing from monolayer molybdenum ditelluride integrated with a silicon nanobeam cavity, 2017, *Nat. Nanotechnol.*, 12, 987, https://doi.org/10.1038/nnano.2017.128

[14] Palacios-Berraquero, C., Kara, D. M., Montblanch, A. R. P., Barbone, M., Latawiec, P., Yoon, D., Ott, A. K., et al., Large-scale quantum-emitter arrays in atomically thin semiconductors, 2017, *Nat. Commun.*, 8, 15093, https://doi.org/10.1038/ncomms15093

[15] Wu, S., Buckley, S., Schaibley, J. R., Feng, L., Yan, J., Mandrus, D. G., Hatami, F., et al., Monolayer semiconductor nanocavity lasers with ultralow thresholds, 2015, *Nature*, 520, 69, https://doi.org/10.1038/nature14290

[16] Li, M., Li, Q., Brongersma, M. L., Atwater, H. A., Optical devices as thin as atoms., 2024, *Science*, 386, 1226, https://doi.org/10.1126/science.adk7707

[17] Han, Z., Wang, F., Sun, J., Wang, X., Tang, Z., Recent Advances in Ultrathin Chiral Metasurfaces by Twisted Stacking, 2023, *Adv. Mater.*, 35, 2206141, https://doi.org/10.1002/adma.202206141

[18] Meng, Y., Feng, J., Han, S., Xu, Z., Mao, W., Zhang, T., Kim, J. S., et al., Photonic van der Waals integration from 2D materials to 3D nanomembranes, 2023, *Nat. Rev. Mater.*, 8, 498, https://doi.org/10.1038/s41578-023-00558-w

[19] Choo, S., Varshney, S., Liu, H., Sharma, S., James, R. D., Jalan, B., From oxide epitaxy to freestanding membranes: Opportunities and challenges, 2024, *Sci. Adv.*, 10, 8561, https://doi.org/10.1126/sciadv.adq8561

[20] Rogers, J. A., Lagally, M. G., Nuzzo, R. G., Synthesis, assembly and applications of semiconductor nanomembranes, 2011, *Nature*, 477, 45, https://doi.org/10.1038/nature10381

[21] Celano, U., Schmidt, D., Beitia, C., Orji, G., Davydov, A. V., Obeng, Y., Metrology for 2D materials: a perspective review from the international roadmap for devices and systems, 2024, *Nanoscale Adv.*, 6, 2260, https://doi.org/10.1039/D3NA01148H

[22] Hsu, C. W., Zhen, B., Stone, A. D., Joannopoulos, J. D., Soljacic, M., Bound states in the continuum, 2016, *Nat. Rev. Mater.*, 1, 16048, https://doi.org/10.1038/natrevmats.2016.48

[23] Koshelev, K., Lepeshov, S., Liu, M., Bogdanov, A., Kivshar, Y., Asymmetric Metasurfaces with High-Q Resonances Governed by Bound States in the Continuum, 2018, *Phys. Rev. Lett.*, 121, 193903, https://doi.org/10.1103/PhysRevLett.121.193903

[24] Richter, F. U., Sinev, I., Zhou, S., Leitis, A., Oh, S. H., Tseng, M. L., Kivshar, Y., et al., Gradient High-Q Dielectric Metasurfaces for Broadband Sensing and Control of Vibrational Light-Matter Coupling, 2024, *Adv. Mater.*, 36, 2314279, https://doi.org/10.1002/adma.202314279

[25] Santiago-Cruz, T., Gennaro, S. D., Mitrofanov, O., Addamane, S., Reno, J., Brener, I., Chekhova, M. V., Resonant metasurfaces for generating complex quantum states, 2022, *Science*, 377, 991, https://doi.org/10.1126/science.abq8684

[26] Yang, J. H., Huang, Z. T., Maksimov, D. N., Pankin, P. S., Timofeev, I. V., Hong, K. B., Li, H., et al., Low-Threshold Bound State in the Continuum Lasers in Hybrid Lattice Resonance Metasurfaces, 2021, *Laser Photonics Rev.*, 15, 2100118, https://doi.org/10.1002/lpor.202100118

[27] Kühner, L., Sortino, L., Tilmann, B., Weber, T., Watanabe, K., Taniguchi, T., Maier, S. A., Tittl, A., High-Q Nanophotonics over the Full Visible Spectrum Enabled by Hexagonal Boron Nitride Metasurfaces, 2023, *Adv. Mater.*, 35, 2209688, https://doi.org/10.1002/adma.202209688

[28] Liu, C., Hsiao, H., Chang, Y., Nonlinear two-photon pumped vortex lasing based on quasi-bound states in the continuum from perovskite metasurface, 2023, *Sci. Adv.*, 6649, 1, https://doi.org/10.1126/sciadv.adf6649

[29] Romano, S., Mangini, M., Penzo, E., Cabrini, S., De Luca, A. C., Rendina, I., Mocella, V., Zito, G., Ultrasensitive surface refractive index imaging based on quasi-bound states in the continuum, 2020, *ACS Nano*, 14, 15417, https://doi.org/10.1021/acsnano.0c06050

[30] Huang, C., Zhang, C., Xiao, S., Wang, Y., Fan, Y., Liu, Y., Zhang, N., et al., Ultrafast control of vortex microlasers, 2020, *Science*, 367, 1018, https://doi.org/10.1126/science.aba4597

[31] Wu, M., Ha, S. T., Shendre, S., Durmusoglu, E. G., Koh, W. K., Abujetas, D. R., Sánchez-Gil, J. A., et al., Room-Temperature Lasing in Colloidal Nanoplatelets via Mie-Resonant Bound States in the Continuum, 2020, *Nano Lett.*, 20, 6005, https://doi.org/10.1021/acs.nanolett.0c01975

[32] Yesilkoy, F., Arvelo, E. R., Jahani, Y., Liu, M., Tittl, A., Cevher, V., Kivshar, Y., Altug, H., Ultrasensitive hyperspectral imaging and biodetection enabled by dielectric metasurfaces, 2019, *Nat. Photonics*, 13, 390, https://doi.org/10.1038/s41566-019-0394-6

[33] Al-Ani, I. A. M., As'Ham, K., Huang, L., Miroshnichenko, A. E., Hattori, H. T., Enhanced Strong Coupling of TMDC Monolayers by Bound State in the Continuum, 2021, *Laser Photonics Rev.*, 15, 2100240, https://doi.org/10.1002/lpor.202100240

[34] Xu, L., Zangeneh Kamali, K., Huang, L., Rahmani, M., Smirnov, A., Camacho-Morales, R., Ma, Y., et al., Dynamic Nonlinear Image Tuning through Magnetic Dipole Quasi-BIC Ultrathin Resonators, 2019, *Adv. Sci.*, 6, 1802119, https://doi.org/10.1002/advs.201802119

[35] Sadrieva, Z. F., Sinev, I. S., Koshelev, K. L., Samusev, A., Iorsh, I. V., Takayama, O., Malureanu, R., et al., Transition from Optical Bound States in the Continuum to Leaky Resonances: Role of Substrate and Roughness, 2017, *ACS Photonics*, 4, 723, https://doi.org/10.1021/acsphotonics.6b00860

[36] Hu, H., Lu, W., Antonov, A., Berté, R., Maier, S. A., Tittl, A., Environmental permittivity-asymmetric BIC metasurfaces with electrical reconfigurability, 2024, *Nat. Commun.*, 15, 7050, https://doi.org/10.1038/s41467-024-51340-7

[37] Barth, I., Deckart, M., Conteduca, D., Arruda, G. S., Hayran, Z., Pasko, S., Krotkus, S., et al., Lasing from a Large-Area 2D Material Enabled by a Dual-Resonance Metasurface, 2024, *ACS Nano*, 18, 12897, https://doi.org/10.1021/acsnano.4c00547

[38] Konishi, K., Akai, D., Mita, Y., Ishida, M., Yumoto, J., Gonokami, M. K., Circularly polarized vacuum ultraviolet coherent light generation using a square lattice photonic crystal nanomembrane, 2020, *Optica*, 7, 855, https://doi.org/10.1364/OPTICA.393816

[39] Rosas, S., Adi, W., Beisenova, A., Biswas, S. K., Kuruoglu, F., Mei, H., Kats, M. A., et al., Enhanced biochemical sensing with high-Q transmission resonances in free-standing membrane metasurfaces, 2025, *Optica*, 12, 178, https://doi.org/10.1364/OPTICA.549393

[40] Shen, S. J., Lee, B. R., Peng, Y. C., Wang, Y. J., Huang, Y. W., Kivshar, Y., Tseng, M. L., Dielectric High-Q Metasurfaces for Surface-Enhanced Deep-UV Absorption and Chiral Photonics, 2024, *ACS Photonics*, in press, https://doi.org/10.1021/acsphotonics.4c01960

[41] Adi, W., Rosas, S., Beisenova, A., Biswas, S. K., Mei, H., Czaplewski, D. A., Yesilkoy, F., Trapping light in air with membrane metasurfaces for vibrational strong coupling, 2024, *Nat. Commun.*, 15, 10049, https://doi.org/10.1038/s41467-024-54284-0

[42] Ho, Y. L., Fong, C. F., Wu, Y. J., Konishi, K., Deng, C. Z., Fu, J. H., Kato, Y. K., et al., Finite-Area Membrane Metasurfaces for Enhancing Light-Matter Coupling in Monolayer Transition Metal Dichalcogenides, 2024, *ACS Nano*, 18, 24173, https://doi.org/10.1021/acsnano.4c05560

[43] Contractor, R., Noh, W., Redjem, W., Qarony, W., Martin, E., Dhuey, S., Schwartzberg, A., et al., Scalable single-mode surface-emitting laser via open-Dirac singularities, 2022, *Nature*, 608, 692, https://doi.org/10.1038/s41586-022-05021-4

[44] Ma, X., Kudtarkar, K., Chen, Y., Cunha, P., Ma, Y., Watanabe, K., Taniguchi, T., et al., Coherent momentum control of forbidden excitons, 2022, *Nat. Commun.*, 13, 6916, https://doi.org/10.1038/s41467-022-34740-5

[45] Ren, Y., Li, P., Liu, Z., Chen, Z., Chen, Y.-L., Peng, C., Liu, J., Low-threshold nanolasers based on miniaturized bound states in the continuum, 2022, *Sci. Adv.*, 8, 8817, https://doi.org/10.1126/sciadv.ade8817

[46] Kodigala, A., Lepetit, T., Gu, Q., Bahari, B., Fainman, Y., Kanté, B., Lasing action from photonic bound states in continuum, 2017, *Nature*, 541, 196, https://doi.org/10.1038/nature20799

[47] Hwang, M. S., Lee, H. C., Kim, K. H., Jeong, K. Y., Kwon, S. H., Koshelev, K., Kivshar, Y., Park, H. G., Ultralow-threshold laser using super-bound states in the continuum, 2021, *Nat. Commun.*, 12, 4135, https://doi.org/10.1038/s41467-021-24502-0

[48] Gölz, T., Baù, E., Aigner, A., Mancini, A., Barkey, M., Keilmann, F., Maier, S., Tittl, A., Revealing Mode Formation in Quasi-Bound States in the Continuum Metasurfaces via Near-Field Optical Microscopy, 2024, *Adv. Mater.*, 36, 2305978, https://doi.org/10.1002/adma.202405978

[49] Hsu, C., Frisenda, R., Schmidt, R., Arora, A., de Vasconcellos, S. M. Bratschitsch, R., van der Zant, H. S. J., Castellanos-Gomez, A., Thickness-dependent refractive index of 1L, 2L, and 3L $MoS_2$, $MoSe_2$, $WS_2$, and $WSe_2$, 2019, *Adv. Optical Mater.*, 7, 1900239, https://doi.org/10.1002/adom.201900239

[50] Beliaev, L. Y., Shkondin, E., Lavrinenko, A. V., Takayama, O., Optical, structural and composition properties of silicon nitride films deposited by reactive radio-frequency sputtering, low pressure and plasma-enhanced chemical vapor deposition, 2022, *Thin Solid Films*, 763, 139568, https://doi.org/10.1016/j.tsf.2022.139568

**Supporting Information**

Supporting Information is available from the Wiley Online Library.